# Power Law Distribution of the Duration and Magnitude of Recessions in Capitalist Economies : Breakdown of Scaling

Paul Ormerod (Pormerod@volterra.co.uk) *

and

Craig Mounfield (Craig.Mounfield@volterra.co.uk)

Volterra Consulting Ltd

The Old Power Station
121 Mortlake High Street
London SW14 8SN

13th November 2000

* Corresponding Author

# Abstract

*Power law distributions of macroscopic observables are ubiquitous in both the natural and social sciences. They are indicative of correlated, cooperative phenomena between groups of interacting agents at the microscopic level. In this paper we argue that when one is considering aggregate macroeconomic data (annual growth rates in real per capita GDP in the seventeen leading capitalist economies from 1870 through to 1994) the magnitude and duration of recessions over the business cycle do indeed follow power law like behaviour for a significant proportion of the data (demonstrating the existence of cooperative phenomena amongst economic agents). Crucially, however, there are systematic deviations from this behaviour when one considers the frequency of occurrence of large recessions. Under these circumstances the power law scaling breaks down. It is argued that it is the adaptive behaviour of the agents (their ability to recognise the changing economic environment) which modifies their cooperative behaviour.*

## 1. Introduction

The analytical tools used to characterise the behaviour of complex, disordered systems such as spin glasses [1] has become a paradigm for the study of a wide variety of other systems. These include biological, social and economic systems [2]. In the context of economic and social systems, the behaviour of the system is determined by the interaction of a large number of heterogeneous agents. Each of these agents has their own objectives (typically the maximization of their individual utilities) and has access only to a restricted subset of all the possible available information (thus each agent makes decisions based upon bounded rationality). The heterogeneous nature of the agents gives rise to a dynamical evolution of the system characterised by interactions which are both frustrated and disordered. It is not surprising therefore that power laws may characterise phenomena such as the distribution of wealth [3] of individuals in a social economy.

It is commonly assumed that the observation of power-law distributions (fractal behaviour) in a systems macroscopically observable quantities is a characteristic *footprint* of cooperative behaviour in many-body systems representing the effects of correlated behaviour amongst the systems constituents. Power law distributions are both self-similar and scale free demonstrating that events may occur on all length and time scales. It is also argued that the precise details of the microscopic interactions between the systems constituents do not alter the fundamental nature of the distribution of macroscopic events observed i.e. the phenomenon of universality.

There is a crucial difference however between 'dumb' particles interacting with one another via fixed physical laws and economic agents (individuals) interacting with one another; namely that individuals are capable of adapting their rules of interaction to reflect the economic environment in which they are placed.

In this paper we will argue that when one is considering aggregate macroeconomic data (annual growth rates in real per capita GDP in the seventeen leading capitalist economies from 1870 through to 1994) the magnitude and duration of recessions over the business cycle do indeed follow power law like behaviour for a significant proportion of the data (demonstrating the existence of cooperative phenomena amongst the economic agents). Crucially, however, there are systematic deviations from this behaviour when one considers the frequency of occurrence of large recessions. Under these circumstances the power law scaling breaks down. It is argued that it is the adaptive behaviour of the agents (their ability to recognise the changing economic environment) which modifies their cooperative behaviour.

## 2. The Data

The data on real per capita GDP in dollars is taken from the OECD monograph [4]. Data for each year from 1870 through 1994 is used for the following countries: Australia, Austria, Belgium, Canada, Denmark, Finland, France, Germany, Italy, Japan, Netherlands, Norway, New Zealand, Sweden, Switzerland, United Kingdom and the United States.[1] Details of the method of calculating an appropriate set of exchange rates with which to convert all the series to a real dollar basis, and also the procedures to allow for variations in the size of the territory of some of the countries over time may be found in[4].

Annual growth rates are calculated by taking the percentage change of the data. In much applied economic analysis, growth rates are defined as the first difference of the natural logs of the data. The results are very similar regardless of which of the two definitions is used. But the former is easier to appreciate in this context, given that some very large falls in output have taken place so the two calculations do not yield almost identical numbers. For example, in 1945 output in Japan fell by 49 per cent, a figure which can perhaps be grasped more readily than the fact that the natural log of Japanese output fell by 0.68 in that year.

---

[1] Data on Japan are only available from 1885, and for Switzerland from 1900

**2.1    The Duration of Recessions**

The first step in the analysis was to focus on the periods of recession, defined as being those years in which growth was less than zero.  Although there is debate as to what growth rate technically constitutes a recession, this measure is unequivocal.  The duration of any recession was then calculated for each country, allocated into the appropriate group and the totals for each group added up.

In other words, the number of times a recession lasted just one year for any given country was calculated, the number of times a recession lasted two years, and so on.  Recessions lasting more than one year are only allocated to one category, so that, for example, a period in which growth was negative for three successive years is counted as a three year recession, and the years in this sequence are not allocated as well into the one and two-year categories.  The longest recession noted was in Austria, when growth was negative in each of the years from 1913 through 1919, a period of seven years.

The individual country data on duration was pooled, giving a total number of 336 observations across the seventeen country sample.  The distribution is shown in Figure 1, where it can be seen that there were some 200 examples of recessions lasting only one year, just under 100 lasting two years, and so on.

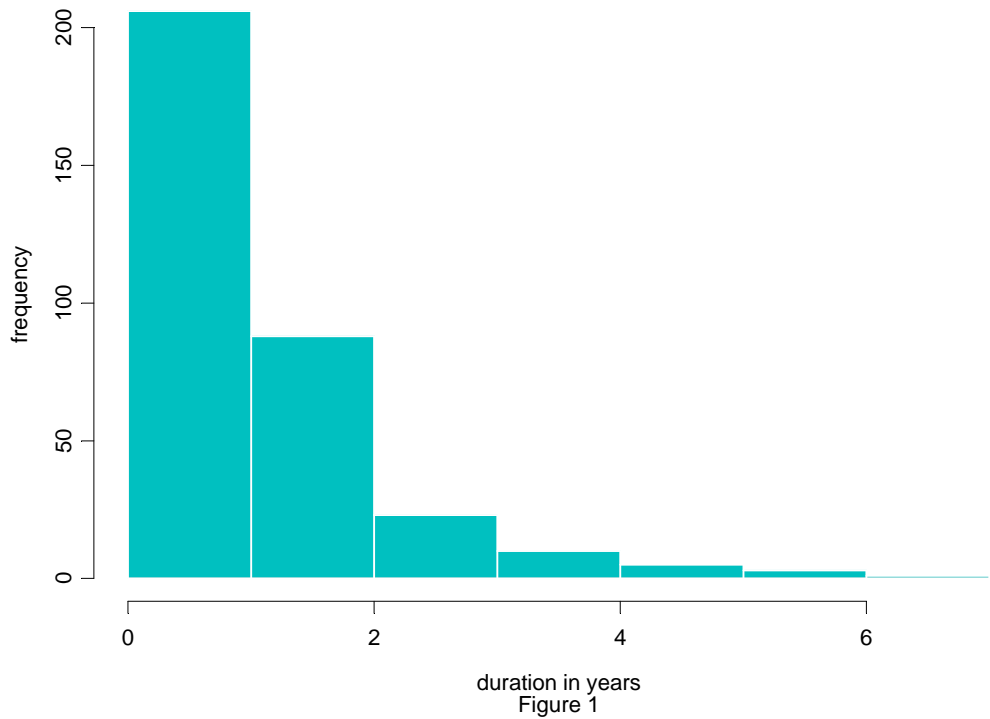

Figure 1

The relationship between the frequency, $N$, of the observations and the duration, $D$, in years of a recession is postulated to be of the form of a power law:

$$N = \frac{\alpha}{D^{\beta}} \qquad (1)$$

Equation (1) could be estimated by either taking a log-log transformation, or by a direct non-linear algorithm. The latter gives a somewhat better fit, and these are the results reported in table 1.

**Table 1    Estimation of equation (2):**

$\alpha$ = 209.6, with a standard error of 13.4

$\beta$ = -1.69, with a standard error of 0.20

standard error of the equation is 13.54 (28.2 per cent of the mean of the dependent variable).

The robustness of these results was checked in two ways. First, equation (1) was estimated for the eight individual countries where a recession of at least five years' duration had taken place. Qualitatively, the results are identical to those reported above for the pooled sample of 17 countries. The exponent is not significantly different in any case from the -1.69 of the pooled sample.

Similar results were also obtained using two different definitions of recession years. The overall mean growth rate of each country was subtracted from the actual growth rate, and observations of this variable of less than zero were classified as recession years. The second definition was a refinement of this concept. Studies have identified four distinct regimes of capitalist economies, from 1870-1913, 1913-50, 1950-73 and 1973 onwards [4]. The sample mean of each country during each of these regimes was subtracted from the actual growth rate, and again observations of less than zero were counted as recession years. The exponent on the duration variable in the regressions was, respectively, -1.36 and -1.28, similar to the -1.69 of the pooled sample.

Figure 2 plots the actual and fitted values of (1). With a perfect fit, all the points in the chart would lie on the 45 degree line.

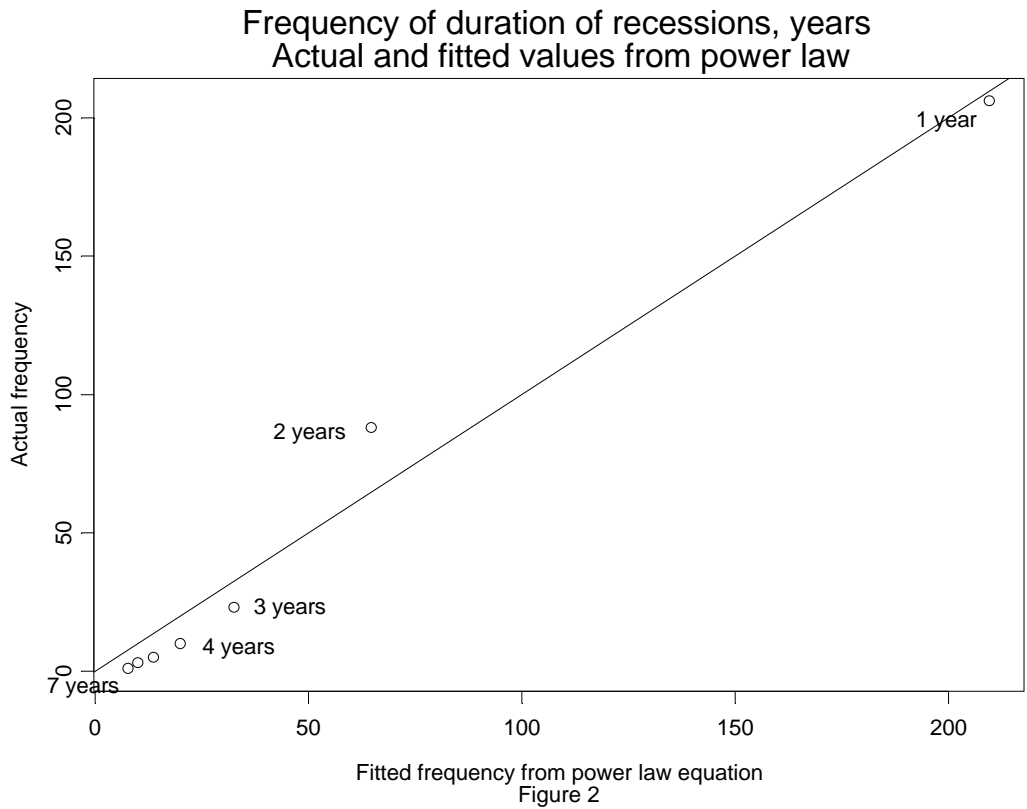

Figure 2

The fit is not unreasonable. However, it is clear that the power law relationships suggests that there should have been more recessions of long duration than have actually existed.

| **Duration (years)** | 1 | 2 | 3 | 4 | 5 | 6 | 7 |
|---|---|---|---|---|---|---|---|
| Actual | 206 | 88 | 23 | 10 | 5 | 3 | 1 |
| Fitted | 210 | 65 | 33 | 20 | 14 | 10 | 8 |

It is not really surprising that a power law over-predicts the number of long recessions in capitalist economies. Unlike grains of sand in experiments to measure the size of avalanches, individual agents in economies, whether governments, firms, or individuals, do have the ability to adapt their behaviour.

Inspecting Figure 2 more closely, the reason for the over-prediction of the longer duration recessions becomes clearer. There appear to be two distinct factors at work. The observations for durations of between two and seven years almost form a straight line, with the one-year recession observation being a distinct outlier.

This is confirmed by a formal regression using the data for (1) omitting the one year duration observation. The exponent on the duration variable is -3.23 with a standard error of only 0.05 and the fit is extremely good, with the equation standard error being only 0.3 per cent of the mean of the dependent variable. Figure 3 plots this.

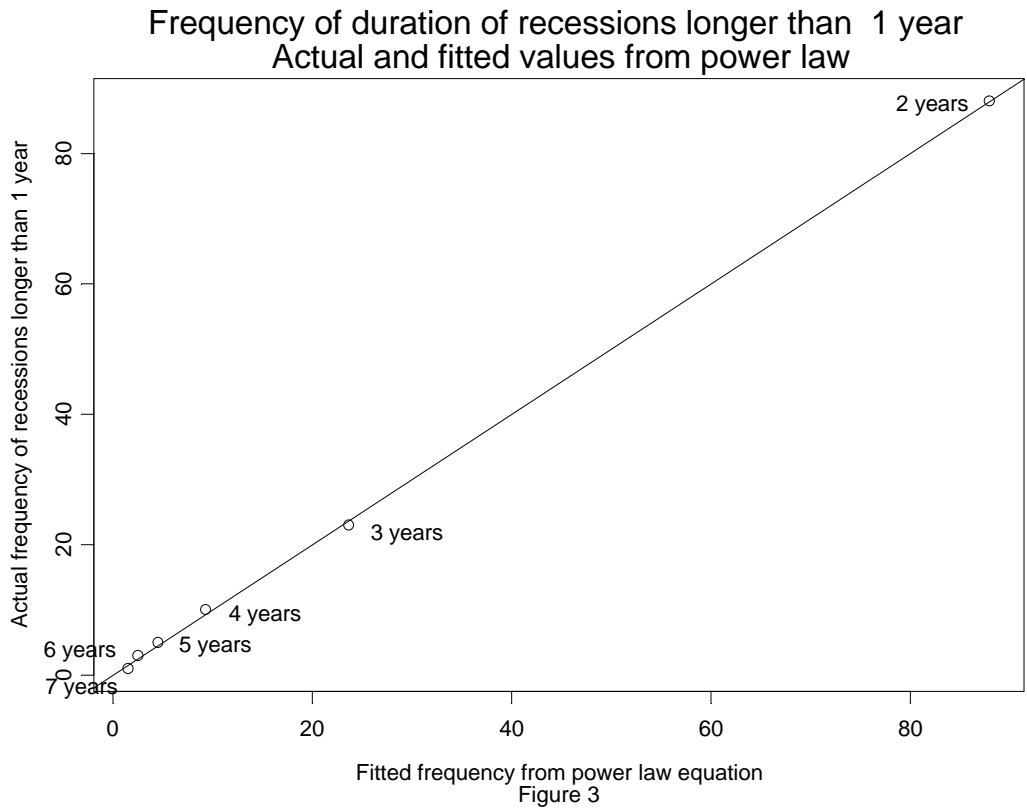

Figure 3

**Table 3. Duration of recessions greater than 1 year: Actual and fitted values from power law**

| Duration (years) | 2 | 3 | 4 | 5 | 6 | 7 |
|---|---|---|---|---|---|---|
| Actual | 88 | 23 | 10 | 5 | 3 | 1 |
| Fitted | 88 | 24 | 9 | 5 | 3 | 2 |

A similar finding is obtained using the two alternative definitions of recession years discussed above. The exponent is again larger in absolute value than when the one year observation is included, and the fit is much better. Using any of these regressions to extrapolate the predicted number of one year recessions gives a larger figure than that actually observed. Using the zero growth definition, for example, 256 examples of one year recessions are predicted, compared to the actual number of 208.

This suggests that there may be two separate processes going on in the process which generates data on capitalist recessions. When a recession arises, for whatever reason, agents appear to have some capacity to react quickly, which often prevents the recession from being prolonged beyond one year. Once this has happened, however, recessions can take place on all scales of duration, exactly as in the sand pile experiment.

Keynes [5] emphasised the importance of expectations in recessions in the capitalist economies. He suggested that once expectations, particularly those of companies, about the future become particularly depressed, they are then very hard to revive. In these circumstances, companies, far from adapting in a positive manner, continue to cut back on investment and employment, thereby further prolonging the recession. The empirical evidence is certainly consistent with this view.

## 2.2 The Size of Recessions

We measure the size of a recession as the cumulative fall in output during the recession. So, for example, in the four-year recession in the United States in the early 1930s, output per head fell by a cumulative total of 34.8 per cent.

An important point to note is that the correlation between the duration of a recession and its cumulative size is not strong. Over the sample as a whole, the Pearson sample correlation coefficient between duration and size is only 0.52 (the absolute value of the size is used, so that a positive correlation indicates that longer durations tend to be associated with large falls in output). Trimming the most extreme outliers in the size data, mainly because enormous falls in output were recorded over short periods of time in the defeated and some of the occupied countries at the end of the world wars, makes little difference to the result. The strongest correlation is obtained by omitting 20 per cent of the sample, which gives a point estimate of 0.63.

The basic data for the cumulative size of recessions is plotted in figure 4, which groups the data into bands of 2 percentage points.

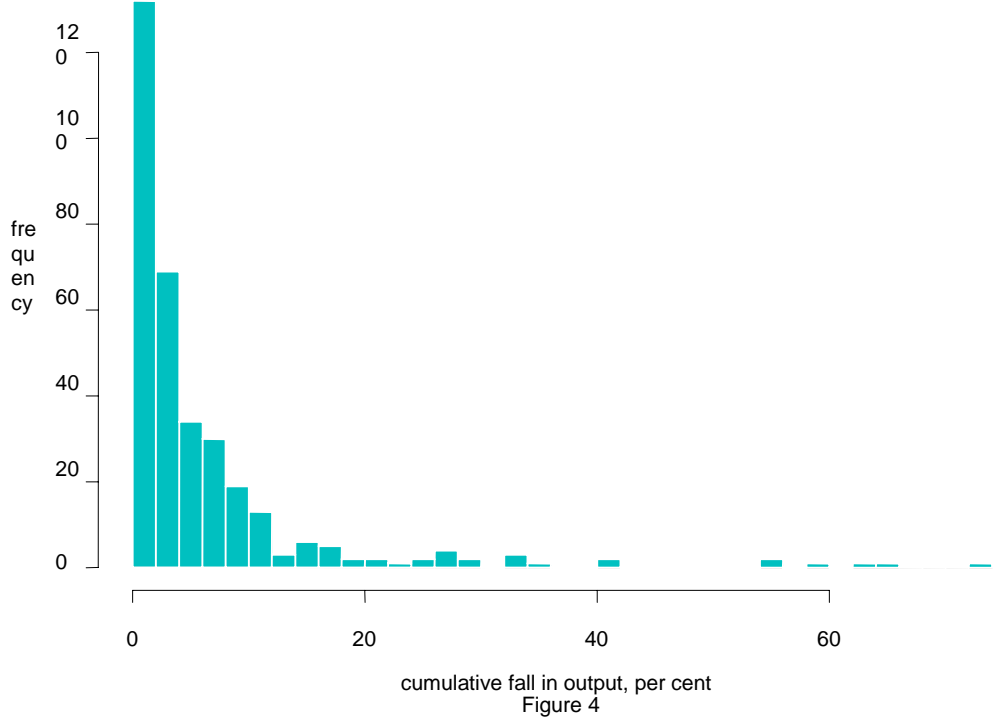

Figure 4

The data in Figure 4 are grouped into intervals each spanning two percentage point, giving 132 observations for which the cumulative fall in output was between 0 and 2 per cent, 69 when it was between 2 and 4 per cent, 34 between 4 and 6 per cent, and so on. There are eight observations where the fall is over 40 per cent: these are Austria, France, Germany, Italy, Japan and the Netherlands during the Second World War, Austria during the First World War, and Canada during the Great Depression of the 1930s.

A power law produces an approximation to this data, with the exponent in the non-linear regression being -1.29 with a standard error of 0.045. But the equation standard error is 41.3 per cent of the mean of the dependent variable. The essential problem is that the estimated equation predicts considerably too many very large recessions, as Figure 5a shows.

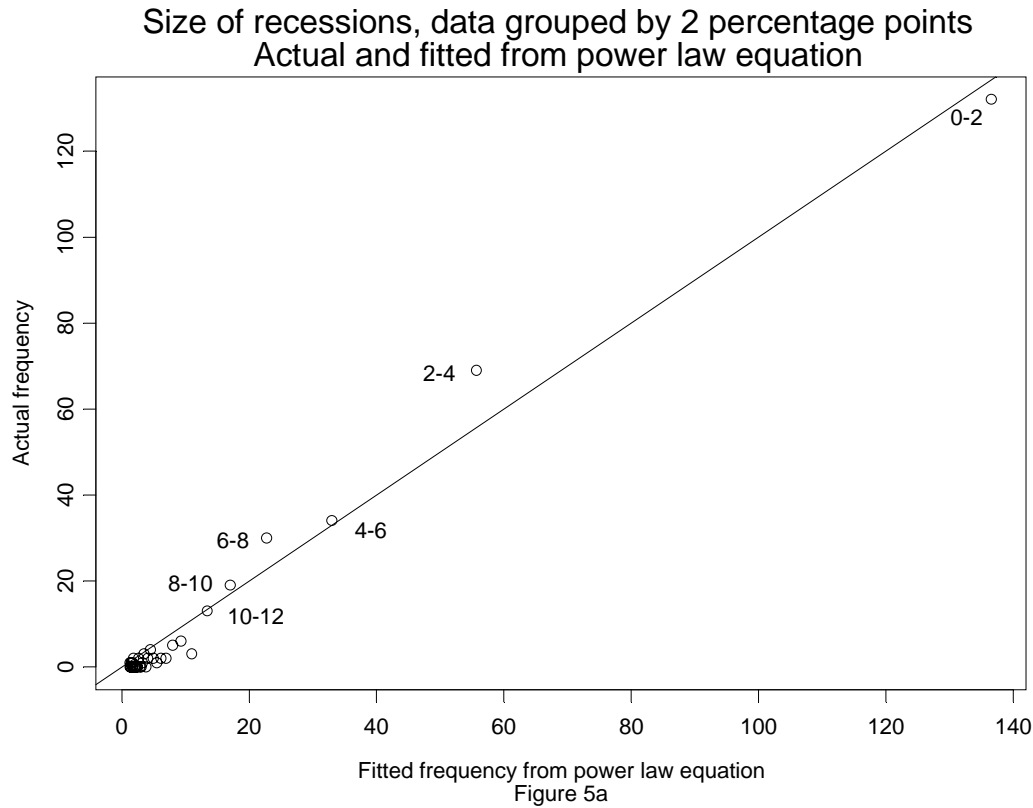

Figure 5a

Omitting the seven largest values from the sample, which are all associated with world wars, improves the fit, with the equation standard error being 30.3 per cent of that of the dependent variable. The exponent is -1.28 with a standard error of 0.06. However, the problem of over-prediction of large recessions remains.

With the duration data, omitting durations of just one year led to a very marked improvement in fit of a power law relationship to the remaining data. Various experiments were carried out on the size data, leaving out durations of just one year, leaving out the smallest size recessions, and so on. But the essential problem of over-prediction of the tail of the distribution was not solved.

Grouping the data into wider bands gives somewhat better results. There have been, for example, 201 recessions in which the cumulative fall in output was between 0 and 4 per cent, 64 in which it was between 4 and 8 per cent, and so on. Using the 4 percentage point bands gives an equation with an estimated exponent of -1.81 and a standard error of 0.054. The equation standard error is 18.6 per cent of the mean value of the dependent variable. The actual and fitted values for this equation are plotted in Figure 5b.

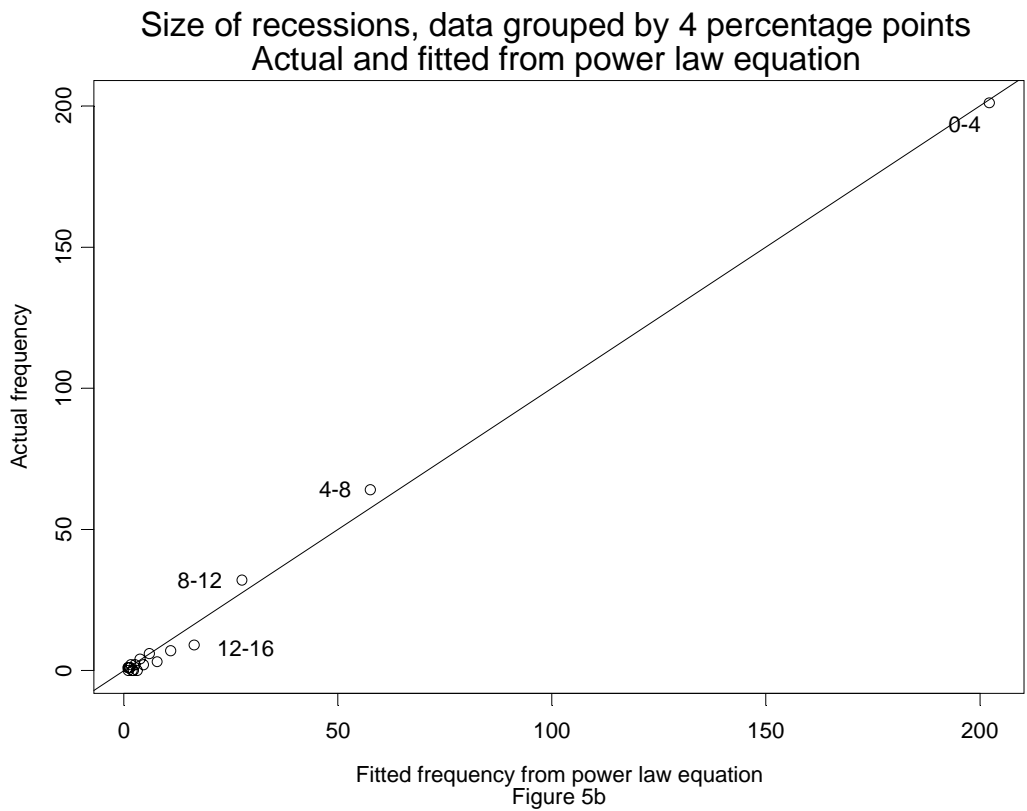

Figure 5b

Table 3, however, shows that even here the over-fitting of the extreme values remains.

**Table 3. Cumulative size of recessions, actual and fitted from power law**

| Size (%) | Actual number | Fitted |
|---|---|---|
| 0-4 | 201 | 202 |
| 4-8 | 64 | 58 |
| 8-12 | 32 | 28 |
| 12-16 | 9 | 16 |
| 16-20 | 7 | 11 |
| 20-24 | 3 | 8 |
| 24-28 | 6 | 6 |
| 28-32 | 2 | 5 |
| 32-36 | 4 | 4 |
| 36+ | 8 | 17 |

## 3. Conclusion

In this paper, we have examined the annual real GDP growth rate data for seventeen capitalist economies over the period 1870 - 1994. The purpose was to investigate to what extent the structure of cyclical recessions and expansions could be approximated by a power law distribution of events.

It has been found that power law distributions do provide an approximation to both the durations and sizes of recessions under capitalism. In each case, however, there is an important qualification: estimated power law relationships suggest that there should have been more large recessions, both in terms of duration and of size, than has actually been the case. This suggest that agents do have a certain amount of adaptability once a recession starts, which sometimes prevents it from becoming large or for lasting longer than a year.

The exception to this qualification is for recessions with a duration longer than a year, when a power law fits the duration data almost perfectly. This implies that once a recession has lasted for more than a year then there exists the possibility of the recession lasting for any duration. In economic terms we might think of the long-term expectations of agents, a concept emphasised by Keynes in his work on the Great Depression of the 1930s. Once the long-term expectations of agents become depressed, they appear to be very hard to revive.